\documentstyle[12pt,epsf]{article}
\def\beq{\begin{equation}}
\def\eeq{\end{equation}}
\def\bea{\begin{eqnarray}}
\def\eea{\end{eqnarray}}
\def\R{\rangle}
\def\L{\langle}
\def\lt{\left}
\def\rt{\right}
\newcommand{\sectiono}[1]{\section{#1}\setcounter{equation}{0}}

\begin{document}
{}~
\hfill\vbox{\hbox{hep-th/0101014}\hbox{MRI-P-001203}
}\break

\vskip 1.4cm

\centerline{\large \bf Test of Siegel Gauge for the Lump Solution}

\vspace*{5.0ex}
\centerline{\large \rm Partha Mukhopadhyay and 
Ashoke Sen}

\vspace*{6.5ex}

\centerline{\large \it Harish-Chandra Research
Institute\footnote{Formerly Mehta Research Institute of Mathematics
and Mathematical Physics}}

\centerline{\large \it  Chhatnag Road, Jhusi,
Allahabad 211019, INDIA}
        
\centerline{E-mail: partha@mri.ernet.in, asen@thwgs.cern.ch,
sen@mri.ernet.in
}

\vspace*{8.5ex}

\centerline{\bf Abstract}
\bigskip

We test the validity of the Siegel gauge condition for the lump solution
of cubic
open bosonic string field theory by checking the equations of motion of
the string field components outside the Siegel gauge. At level (3,6)
approximation, the
linear and quadratic terms of the equations of motion of these fields are
found to cancel within about 20\%.  

\vfill \eject

\baselineskip=17.3pt

\tableofcontents

\sectiono{Introduction} \label{s1}

String field theory\cite{W,BER} 
has turned out to be a very powerful
tool
for
directly 
verifying the various conjectures\cite{conj,WK,HO} about
tachyon
condensation on unstable D-branes of bosonic and superstring theories.
There are two main conjectures:
1) the state corresponding to the tachyon condensed
to the minimum of its potential is the closed string vacuum
state without any D-brane, and 2) suitable classical 
solutions
involving the tachyon field
represent the various lower dimensional D-branes.
Both the conjectures 
make statements about nonperturbative field configurations on a D-brane. It is
natural that in a second quantized string theory which defines string theory 
off-shell, one should be able to verify these conjectures directly.
Various works in second 
quantized string theories have been done towards this direction both in
the context of bosonic string
\cite{SZ,T,MT,HK,MJMT,MSZ,MR,M,SZ2,T2,0008252,HS,0010190,0011238,
0012251,GS,KMM,0009191,MN}
and
superstring theories
\cite{B,BSZ,SR,IN,KMM2}. 

Witten's cubic string field theory\cite{W} is one of the candidates for the 
second quantized string theory in the bosonic case. Although the action of this
theory is at most cubic in fields, there are infinite number of terms in 
the action. Therefore performing computations is in general difficult. But 
fortunately the argument of universality\cite{univ,RZ} and  
application of symmetry properties, $-$ all of which help in truncating
the
string
field 
consistently to a subspace of the full configuration space, $-$ combined
with 
the
level truncation method 
introduced by Kostelecky and Samuel\cite{KS} have made computations 
possible. The work of refs.\cite{SZ,MT} has verified the first conjecture 
in bosonic string theory to a very high degree of accuracy. In the work of
ref.\cite{MSZ} the verification of the 
second conjecture has been performed by explicitly 
constructing the codimension one lump solution on a D-brane wrapped on a
circle, and verifying that the energy of the solution 
corresponds to that of the unwrapped D-brane of one lower dimension.
Computations
have been done for various values of the radius of the circle and with a 
modified level truncation scheme. This analysis has also been extended to 
the higher codimension solutions\cite{MR,M}.
   
The cubic string field theory is a gauge theory. Therefore in this case one
 has
to deal with the gauge-fixing conditions. Because of the infinite number
of fields present in the theory,  it is difficult to show the
reasonability of a given gauge choice in
general. Both, the construction of  the nonperturbative vacuum  
solution\cite{SZ,MT} 
and the lump solution\cite{MSZ,MR,M} have been carried out using the Siegel
gauge condition. But acceptance of these solutions is subject to the 
validity of this gauge chosen for the solutions.
In ref.\cite{HS} the stringy BRST invariance of the 
nonperturbative vacuum solution has been explicitly checked in the level 
truncation scheme. This is equivalent to checking the validity of the 
Siegel gauge for the solution by ensuring that the equations of motion of 
the fields outside the Siegel gauge are automatically satisfied by the
solution obtained in the Siegel gauge. In this paper we attempt to check
the validity
of the Siegel gauge for the lump solution given in ref.\cite{MSZ} using
similar method.  
We will
perform computations only for the value of the radius 
$\sqrt 3$ and use the modified level truncation method introduced in 
ref.\cite{MSZ}. 
  
The rest of the paper is organized as follows. In sec.\ref{setup} we give
the
description of the setup in which we perform our computation. 
This is basically
a review of the setup considered in ref.\cite{MSZ}, the only exception 
being that the set of string fields is extended to include the fields
outside the Siegel 
gauge. In sec.\ref{validity} we explain, in the context of the lump
solution, the approach taken to test the
validity
of a gauge choice for a specific solution.
Finally in sec.\ref{check-eqn} we give our results up to the level $(3,~6)$ 
in the modified level truncation scheme. Although this is done only for
radius $R=\sqrt 3$, the analysis can be easily extended to the other
values of the radius
analysed in ref.\cite{MSZ}.
%
\sectiono{Review of the Setup}  
\label{setup}

Here we review the general setup considered in ref.\cite{MSZ} in computing the 
lump solution in a modified level truncation scheme. We will follow the same
notations and conventions adopted in this paper.

\begin{itemize}
\item
{\bf Action:} We consider a D-brane in the 26 dimensional bosonic string 
theory. The cubic open 
string field theory action on the D-brane is
given by:
\bea
S = - \frac{1}{g_o^2} \lt( \frac{1}{2} \L \Phi, Q_B \Phi \R + \frac{1}{3}
\L \Phi, \Phi * \Phi \R \rt),
\label{action}
\eea
where the string field $\Phi$ is a ghost number 1 state in the
Hilbert space of the combined matter-ghost conformal field theory. 
The BRST charge $Q_B$, the BPZ inner product
$\langle A,B\rangle$
and the star product $A*B$ have their usual
meaning. This action
possesses gauge invariance with the following gauge transformation law:
\bea
\delta |\Phi \R = Q_B |\Lambda \R + |\Phi * \Lambda \R - |\Lambda * \Phi \R,
\label{gauge.trf}
\eea 
where $|\Lambda \R $ is a ghost number zero state.
\item
{\bf Background:} The background considered in ref. \cite{MSZ} is of quite 
general type. The total matter conformal field theory CFT on the open
string world-sheet is,
\bea
\mbox{CFT} = \mbox{CFT($X$)} + \mbox{CFT}^\prime ,
\eea
with
\bea
\mbox{CFT}^\prime = \mbox{CFT($Y$)} \oplus \mbox{CFT($X^0$)} \oplus 
\mbox{CFT(${\cal M}$)}.
\eea
Here $X$ is the world-sheet scalar field corresponding to a direction
$x$ along the D-brane
which is compactified on a circle of radius $R$ and 
the scalar fields $Y$ and $X^0$ correspond to respectively a
space-like non-compact direction $y$ transverse to the D-brane, and
the time direction $x^0$. 
${\cal M}$ is an arbtrary manifold describing the rest of the compactification
of space-time with the only restriction that any noncompact direction of
${\cal M}$ is transverse to the D-brane\footnote{ This restriction was
made in ref. \cite{MSZ}
just to make the D-brane have a finite mass.}.
This effectively makes the 
D-brane a D-string aligned along $x$ in the non-compact part 
of the space-time. 

\item
{\bf Consistent truncation:} 
For a lump solution which varies only 
along $x$, the background 
fields can carry  momentum only along the
$x$ direction. As was shown in ref.\cite{MSZ},  
in constructing this solution we can use a truncated version of the string
field theory where we take the string field to be a linear combination of
states created by the ghost oscillators, and the Virasoro generators of
CFT($X$) and CFT$'$, 
on parity even primary states of CFT($X$). Furthermore,
the Siegel gauge condition
excludes the excitations involving the ghost oscillator $c_0$. Since in 
the gauge invariant action one has to include these states, the
truncated spectrum
is generated by
acting the oscillators
\beq
\lt\{ L^X_{-1},L^X_{-2},\cdots ; L^\prime_{-2}, L^\prime_{-3},\cdots ;
c_1, c_0, c_{-1}, c_{-2}, \cdots ; b_{-2}, b_{-3}, \cdots \rt\}    
\label{gauge-unfxd-spct}
\eeq
on either
\begin{itemize}
\item
the zero momentum parity-even primary states of CFT($X$) (with the 
null states removed),
or
\item the Fock vacuum states of the form 
$\mbox{cos}\lt( \frac{n}{R} X(0)\rt) |0\R$. 
\end{itemize}

Among these states we must keep only those which have ghost number 1 and
are twist even. The latter property requires that the total contribution
to $L_0$ eigenvalue of the state from the matter and ghost
{\it oscillators} is odd.

\item
{\bf Modified level truncation scheme:} Ref.\cite{MSZ} has 
introduced a modified level truncation scheme which is applicable in a more
general situation. The level of a field is 
defined as the 
difference between the $L_0$ eigenvalue of the corresponding state and
the zero 
momentum tachyon state. Then the level $(M,N)$ approximation to
the action is obtained by keeping all fields with level $M$ and
below, and keeping all terms in the action with total level $N$
and below.

\end{itemize}
\sectiono{Testing Validity of the Siegel Gauge for the Lump Solution} 
\label{validity}

While making a gauge choice it has to be ensured that any field configuration
can be brought to lie on the gauge slice under a gauge tranformation. Whether
a gauge choice is good in this sense or not can depend on the class of field 
configurations considered in a given problem. For example, for the twist
even 
cofigurations, the Siegel gauge, namely $b_0 |\Phi \R = 0$, is a good gauge 
choice near $\Phi = 0$. But the vacuum solution found in
refs.\cite{KS,SZ,MT} or the lump solution found in 
ref.\cite{MSZ} is far away from $\Phi = 0$.  
Thus the arguments showing the validity of the Siegel gauge near $\Phi=0$
are not applicable here.
In order to show that 
the 
Siegel gauge is a valid gauge choice for a solution of this kind, one
needs to
verify that the solution obtained in the Siegel gauge satisfies the full
set of gauge invariant equations of motion.
Since the solutions in the Siegel gauge are constructed by setting to zero
the variation of the 
gauge fixed 
action with respect to the fields {\em satisfying the
Siegel gauge condition}, what needs to be checked is that the equations of
motion 
(derived from the gauge invariant action)  
of the fields which {\em do not satisfy the Siegel gauge condition}
are also satisfied. For the vacuum solution this was verified in ref.\cite{HS}.
To check the validity of Siegel gauge for the lump solution in this approach
one follows the following steps in the modified level truncation scheme:

\begin{itemize}

\item
Compute the action $S$ (eqn. (\ref{action})) by expanding the string field
$|\Phi \R$ in the truncated Hilbert space without implementing
any gauge condition.\footnote{i.e. using the 
oscillators in (\ref{gauge-unfxd-spct})}

\item 
Find the equations of motion of the fields outside the Siegel gauge.

\item 
Check if these equations of motion are satisfied by the lump solution obtained 
in the gauge fixed theory.

\end{itemize}

A given term in the action
$S$ can be linear, quadratic or cubic in a given field.
Since in the lump solution all the fields outside the Siegel 
gauge are set to zero, in verifying the equations of motion we need to
take the first derivative of the action with respect to various fields and
then set the fields outside the Siegel gauge to zero. Thus only those 
terms in the action which are linear in fields outside the Siegel gauge 
contribute to the equation of motion of these fields.
These are the terms we need to compute.

Let us take the following expansion for the truncated string field:
\bea
|\Phi \R = \sum_a \phi_a |\Phi_a \R + \sum_n \phi_n 
|\Phi_n \R, 
\eea
where the indices $a$ and $n$ run over respectively the states inside the
Siegel
gauge and outside the Siegel gauge. Then the part of the action
which contributes to the equation of motion of a  
specific field $\phi_n$ is:
\bea
-\frac{1}{g_o^2} \lt( \sum_{a} C_{an}~\phi_a +  \sum_{a, b}
C_{abn}~\phi_a \phi_b \rt) \phi_n ,
\eea
where  
$C_{\alpha\beta} = C_{\beta\alpha} = \L \Phi_\alpha, Q_B
\Phi_\beta 
\R$ and 
$C_{\alpha\beta\gamma} = 
C_{\alpha\gamma\beta} = \L
\Phi_\alpha,\Phi_\beta 
* \Phi_\gamma \R$.\footnote{That the 
coefficients $C_{\alpha\beta\gamma}$ are same even for non-cyclic
permutations of the 
indices $\alpha,\beta,\gamma$, is a property of the truncated spectrum.
Here the indices $\alpha,\beta,\gamma$ run over all fields
in the truncated spectrum.}  
Therefore the 
correponding equation of motion is:
\bea
\sum_{a} C_{an}~\phi_a + \sum_{a, b} C_{abn}~\phi_a \phi_b  =0.
\label{eqn-of-motion}
\eea
One has to check 
the above equation in the modified level 
truncation scheme.
\sectiono{Checking Equations of Motion for Fields Outside the Siegel Gauge 
}  
\label{check-eqn}

Here we will check the equations of motion for fields outside the Siegel gauge
in the modified level truncation scheme. We will present results upto the 
level $(3,~6)$. 

We start by making a list of the relevant fields. 
In table \ref{t1} some classes of states have been listed with their vertex 
operators and levels. The relevant fields which get involved in the 
computations are the coefficients of these states in the expansion of the 
string field. The states which are inside the Siegel gauge, namely 
$|T_n \R$, $|U_n \R$, $|V_n \R$, $|W_n \R$ and $|Z_n \R$ are precisely the 
ones which have been considered in ref.\cite{MSZ} for constructing the lump 
solution. The states $|R_n \R$ and $|S_n \R$ are outside the Siegel gauge as 
they involve the $c_0$ oscillator.

Table \ref{t2} shows the fields\footnote{Following ref.\cite{MSZ}, we denote a 
field by the lowercase symbol corresponding to the uppercase symbol used for 
the corresponding state.} and their levels which get involved in our 
computation upto level $(3,~6)$ for $R=\sqrt 3$. The fields in the square 
brackets i.e. $r_0$, $r_1$ and $s_1$ are the ones outside the Siegel gauge. We 
will check the equations of motion for these fields. 

\begin{table}[ht]
\begin{center}
\begin{tabular}{|l|l|c|} \hline\hline
State    & Vertex Operator &  Level \\
         &                 & $R^2=3$ \\ \hline\hline
$|T_n \R = c_1 |\frac{n}{R}\R$ &
$c \, \mbox{cos}\lt(\frac{n}{R} X\rt)$ &  
$n^2/R^2$ \\ \hline
$|U_n \R = c_{-1} |\frac{n}{R}\R$ &
$\frac{1}{2}\,\partial^2c\,\mbox{cos}\lt(\frac{n}{R} X\rt)$ & 
$2 + n^2/R^2$   \\ \hline
$|V_n \R = c_1L^X_{-2} |\frac{n}{R}\R$ &
$ T^X\, c\, \mbox{cos}\lt(\frac{n}{R} X\rt)$ &   
$2 + n^2/R^2$  \\ \hline
$|W_n \R = c_1L^\prime_{-2} |\frac{n}{R}\R$ &
$T^\prime\, c\, \mbox{cos}\lt(\frac{n}{R} X\rt)$ &  
$2 + n^2/R^2$     \\ \hline
$|Z_n \R = c_1L^X_{-1}L^X_{-1} |\frac{n}{R}\R$ &
$c\, \partial^2 \Big(\mbox{cos}\lt(\frac{n}{R} X\rt)\Big)$ &  
$2 + n^2/R^2$    \\ \hline
$|R_n \R = b_{-2}c_0c_1|\frac{n}{R}\R$ &
$b\,\partial c\, c\, \mbox{cos}\lt(\frac{n}{R} X\rt)$ &  
$2 + n^2/R^2$    \\ \hline
$|S_n \R = c_0L^X_{-1} |\frac{n}{R}\R$ &
$\partial c \, \partial \Big(\mbox{cos}\lt(\frac{n}{R} X\rt)\Big)$ &   
$2 + n^2/R^2$   \\ \hline
\end{tabular}
\end{center}
\caption{The Hilbert space states relevant for constructing the lump
solution.} \label{t1}
\end{table}

\begin{table}[ht]
\begin{center}
\begin{tabular}{|c|c|} \hline\hline
Level & Fields \\\hline\hline
$0$ & $t_0$ \\\hline  $1/3$ & $t_1$ \\\hline $4/3$ & $t_2$ \\\hline
 $2$ & $u_0,~v_0,~w_0,~[r_0]$ \\ \hline 
$7/3$ & $u_1,~v_1,~w_1,~z_1,~[r_1],~[s_1]$ \\ \hline  
$3$ & $t_3$ \\ \hline  
\end{tabular}
\end{center}
\caption{Fields in the Hilbert space up to level 3.} \label{t2}
\end{table}

The notations that we will use is as follows. Linear part of
eqn. (\ref{eqn-of-motion}), i.e. $\sum_{a} C_{an}~\phi_a$ computed
using the
level
$(M,~2M)$ approximation to the action,   
will be denoted by $L(M,~2M)$.  
Similarly $Q(M,~2M)$ will
denote  
the quadratic part of eqn. (\ref{eqn-of-motion}), namely, 
$\sum_{a, b} C_{abn}~\phi_a \phi_b \,$.

\noindent{\large \bf \underline{Results}}

\medskip

We define $K=\frac{3\sqrt 3}{4}$. For $R=\sqrt 3$ we have the
following
results:

\medskip

\noindent{\bf \underline{The Field $r_0$}:}
\bea
L(2,~4) &=& \frac{1}{2}~v_0  + \frac{25}{2}~w_0 - 3~u_0 \\
Q(2,~4) &=& - K~t_0^2 - \frac{1}{2}~K^{1 - 2/R^2}~t_1^2
+ \frac{\sqrt 3}{2}~t_0 u_0 + \frac{5}{12 \sqrt 3}~t_0 v_0 
+ \frac{125}{12 \sqrt 3}~t_0 w_0 
\eea
Since contribution to the linear term $L$ should come only from a certain
level
which is $(2,~4)$ in this case, $L$ will have the same expression for any 
higher level. $Q$ in general varies as one changes the level.
\bea
Q(7/3,~14/3) &=& Q(2,~4) - \frac{1}{2}~K^{1-8/R^2}~t_2^2 + \frac{\sqrt 3}{4}~ 
K^{-2/R^2}~t_1 u_1 \cr 
\cr
&& + \frac{125}{24 \sqrt 3}~K^{-2/R^2}~t_1 w_1
   + \lt(\frac{8}{27 R^2} + \frac{5}{54} \rt)~K^{1-2/R^2}~t_1 v_1 \cr
\cr
&& + \lt(\frac{11}{8 R^2} - \frac{1}{R^4} \rt)~K^{-1-2/R^2}~t_1 z_1
\eea
\bea
Q(3,~6) &=& Q(7/3,~14/3) + \frac{703}{324 \sqrt 3}~u_0^2  
- \frac{179}{972}~K~v_0^2 - \frac{9475}{972}~K~w_0^2 \cr
\cr  
&&- \frac{5 \sqrt 3}{108}~u_0 v_0 - \frac{125 \sqrt 3}{108}~u_0 w_0 
- \frac{625}{1458}~K~v_0 w_0 
\eea

\medskip

\noindent {\bf \underline{The Field $r_1$}:}

\medskip

Here $L$ gets contribution at the level $(7/3,~14/3)$.
\bea
L(7/3,~14/3) = - \frac{3}{2}~u_1 
+ \frac{1}{2} \lt(\frac{4}{R^2} + \frac{1}{2} \rt)~v_1 + \frac{25}{4}~w_1 
+ \frac{3}{R^2} z_1 
\eea
\bea
Q(7/3,~14/3) &=& -K^{1-2/R^2}~t_0 t_1 - \frac{1}{2}~K^{1-6/R^2}~t_1 t_2
+ \frac{\sqrt 3}{4}~K^{-2/R^2}~t_0 u_1 \cr
\cr
&& + \frac{1}{2} \lt(\frac{1}{R^2} + \frac{5}{16} \rt)~K^{-1 - 2/R^2}~t_0 v_1
+ \frac{125}{24 \sqrt 3}~K^{-2/R^2}~t_0 w_1 \cr
\cr
&& + \lt(\frac{11}{8R^2} - \frac{1}{R^4} \rt) K^{-1 -2/R^2}~t_0 z_1
+ \frac{\sqrt 3}{4} K^{-2/R^2}~t_1 u_0 \cr
\cr
&& + \lt( \frac{5}{54} - \frac{16}{27R^2} \rt) K^{1 - 2/R^2}~t_1 v_0
+ \frac{125}{24 \sqrt 3} K^{-2/R^2}~t_1 w_0
\eea
\bea
Q(3,~6) &=& Q(7/3,~14/3) + \frac{\sqrt 3}{8}~K^{-6/R^2}~t_2 u_1
+ \lt( \frac{5}{64} - \frac{3}{4R^2} \rt)~K^{-1 - 6/R^2}~t_2 v_1 \cr
\cr
&& + \frac{125}{108}~K^{1 - 6/R^2}~t_2 w_1
+ \frac{1}{2} \lt( \frac{11}{8R^2} - \frac{9}{R^4} \rt)~K^{-1 - 6/R^2}~t_2 z_1
\eea

\medskip

\noindent {\bf \underline{The Field $s_1$}:}

\medskip

\bea
L(7/3,~14/3) = \frac{1}{R^2}~u_1 - \frac{3}{R^2}~v_1 
- \frac{2}{R^2}~\lt(1 + \frac{2}{R^2} \rt)~z_1
\eea
\bea
Q(7/3,~14/3) = - \frac{1}{2R^2}~K^{-1 - 2/R^2}~t_0 u_1
+ \frac{1}{2R^2}~K^{-1 - 2/R^2}~t_1 u_0
\eea
\bea
Q(3,~6) &=& Q(7/3,~14/3) + \frac{3}{4R^2}~K^{-1 - 6/R^2}~t_2 u_1
\eea
\begin{table}
\begin{center}\def\st{\vrule height 3ex width 0ex}
\begin{tabular}{|c|c|c|c|} \hline \hline

Field &  $(2, 4)$ &
$(7/3, 14/3)$ & $(3, 6)$
\st\\[1ex] \hline \hline

$t_0$ &  0.25703 & 0.265131 & 0.269224 \st\\[1ex]
\hline

$t_1$ &  -0.384575 & -0.394396 & -0.394969
\st\\[1ex]
\hline

$t_2$ & -0.107424 & -0.12046 & -0.125011 \st\\[1ex]
\hline

$u_0$ & 0.0888087 & 0.0900609 & 0.0969175 \st\\[1ex] \hline 

$v_0$ & -0.00675676 & -0.0175367 & -0.0172906 \st\\[1ex] \hline

$w_0$ & 0.0317837 & 0.0299617 & 0.0320394 \st\\[1ex] \hline

$u_1$ &... & -0.0643958 & -0.0648543 \st\\[1ex] \hline

$v_1$ &... & 0.0540447 & 0.0505836 \st\\[1ex] \hline

$w_1$ &... & -0.0187778 & -0.0189058 \st\\[1ex] \hline

$z_1$ &... & -0.0698363 & -0.0665402 \st\\[1ex] \hline

$t_3$ &... &... & -0.0142169 \st\\[1ex] \hline \hline

\end{tabular}
\end{center}
\caption{The values of various modes of the string field at the stationary
point of the potential for $R=\sqrt{3}$ calculated at
various levels of approximation.}
\label{t3}
\end{table}
\noindent Now to get the numerical values of 
$L$ and $Q$ for different
fields at 
a given level one has to substitute the lump solution obtained in 
ref.\cite{MSZ} at that level. Table \ref{t3} displays these solutions at 
different levels and table \ref{t4} shows the numerical values that we
obtain after 
substituting these solutions in the equations for $L$ and $Q$ given above.

\begin{table}[ht]
\begin{center}\def\st{\vrule height 3ex width 0ex}
\begin{tabular}{|c|l|c|c|c|c|} \hline\hline
Field & Level         &    L       &     Q      &  L+Q   & (L+Q) / L 
\\ \hline\hline
$r_0$ & $(2,~4)$      & 0.127492   & -0.098026   & 0.0294657  & 0.231119 
\st\\[1ex]  \cline{2-6}
      & $(7/3,~14/3)$ & 0.0955702  & -0.0838568 & 0.0117134 & 0.122563 
\st\\[1ex] \cline{2-6}
      & $(3,~6)$      & 0.101095   & -0.0881436 & 0.0129511 & 0.128108
\st\\[1ex] \hline\hline
$r_1$ & $(7/3,~14/3)$ & -0.0410629 & 0.0323025  & -0.00876042 &
0.213342 
\st\\[1ex] \cline{2-6}
      & $(3,~6)$      & -0.0410517 & 0.0325827  & -0.00846901 & 0.206301  
\st\\[1ex] \hline\hline 
$s_1$ & $(7/3,~14/3)$ & 0.00208592 & -0.00198787& 0.0000980474 &
0.0470043 
\st\\[1ex] \cline{2-6}
      & $(3,~6)$      & 0.00173186 & -0.00131896& 0.000412898 & 0.238413 
\st\\[1ex] \hline\hline
\end{tabular}
\end{center}
\caption{The linear ($L$), quadratic ($Q$) and total ($Q+L$) contribution
to the equations of 
motion of the fields outside the Siegel gauge. The last column shows the
degree to which the linear and the quadratic terms cancel.} \label{t4}
\end{table}

{}From table \ref{t4} we see that for the equation of motion of each
field, there is a high degree of cancellation between the linear and
the quadratic terms in the equation of motion. The last column explains
clearly
the degree of this cancellation, $-$ we see that the sum of the quadratic
and the linear term is typically about 10-20\% of the linear term alone.
This may not seem to be a
very good result, but we note that in the corresponding calculation for the
vacuum solution at level (2,6) ($t_0=.544$, $u_0=.190$,  
$v_0=w_0=.0560$)\cite{KS,RZ}, 
the linear and the quadratic terms in the equation of
motion of $r_0$ are given by respectively 0.158 and -0.124. These 
add up to about 22\% of the linear term.  
Thus at this level our results for the lump solution are as good as those
for the vacuum solution.
As
has been verified in ref.\cite{HS}, the cancellation betwen the linear and
quadratic terms in the equations of motion of the vacuum solution 
improves
to about 1\% when we use the solution at level (10,20)
approximation.\footnote{Note that although 
the authors of ref.\cite{HS}  
used the solutions obtained at the level (10,20) approximation,
in computing the contribution to the equations of motion of the
field $r_0$ they only used the level (2,6) approximation. Presumably
contribution from fields at higher level do not alter the results.} 
We could therfore expect a similar improvement of the results for the lump
solution when we go to higher level.

This provides evidence that the equations of motion of the fields outside
the Siegel gauge are automatically satisfied by the solutions obtained in
ref.\cite{MSZ} in the Siegel gauge.

From the last column of table \ref{t4} one may notice that for the field
$s_1$ the cancellation at level $(7/3,~14/3)$ is much better than that at
level $(3,~6)$. This at a first glance, may seem to show that the level
truncation scheme fails to work in this case. This however is not the
case, as can be seen from that fact that the cancellation for the
field $r_1$ at level (7/3, 14/3) is
much worse than that for $s_1$, and so the high degree of cancellation
for the field $s_1$ should be treated as accidental. 
Indeed, at any given level of
approximation it is always
possible to take the independent fields at level 7/3 to be appropriate
linear combinations of $r_1$ and $s_1$ so that the derivative of the
action with respect to one of these fields is very small (or even zero)
when we substitute the 
Siegel gauge solution  obtained at that level. 
It so happens that in the level (7/3, 14/3) approximation
$s_1$ is the linear combination for which the equation of motion is
satisfied very closely. But clearly this does not establish that the level
(7/3, 14/3) approximation is better than the level (3,6) approximation,
$-$ what is required for the approximation to be good is that the
contribution to the equations of motion of both the fields should be
small.  As can be seen from table \ref{t4}, the cancellation for $r_1$ is
actually better at level (3,6) than at level (7/3, 14/3). Thus we cannot
conclude that the cancellation becomes worse when we go from level (7/3,
14/3) to level (3,6). It is true however that on the whole, the degree of 
cancellation does not
improve either in going from level (7/3, 14/3) to level (3,6). This
could be due to the fact that we do not introduce too many new fields in
going from the level (7/3, 14/3) to the level (3,6) approximation.

\medskip

\noindent{\bf Acknowledgement}: We would like to thank B. Zwiebach for 
useful discussions and comments on the manuscript. A.S. would like to
acknowledge the
hospitality of the Theoretical Physics Department of the Tata Institute of
Fundamental Research where part of the work was done.

\end{document}